\documentclass[preprint,pra,showpacs,nofootinbib]{revtex4-1}  % 改用 revtex4-1

\usepackage{amsmath,amssymb,amsthm}
\usepackage[mathscr]{euscript}
\usepackage{graphicx}  % 只加载一次
\usepackage{float}
\usepackage{dcolumn}   % 只加载一次
\usepackage{bm}        % bold math
\usepackage{subfig}    % 替代 subfigure
\usepackage{hyperref}  % 去掉 dvips 选项，让 hyperref 自动检测驱动
\usepackage{xcolor}
\newcommand{\bea}{\begin{eqnarray}}
\newcommand{\eea}{\end{eqnarray}}
\newcommand{\beq}{\begin{equation}}
\newcommand{\eeq}{\end{equation}}

\begin{document}

\title{Optimal estimation of quantum boundary effect in cosmic string space-time}
\author{Yao Jin}
\affiliation{School of Science, Guiyang University, Guiyang, Guizhou 550005, China}

\begin{abstract}
The presence of a cosmic string modifies vacuum fluctuations, making the evolution of a two-level polarizable atom position-dependent. Such modifications produce effects on the atomic dynamics analogous to those induced by a reflecting boundary. We show that these quantum boundary effects can be estimated by performing a sequence of $N$ measurements on a single probe atom. For a fixed total probe time, the precision limit is attained by preparing each probe in its optimal initial state, performing the corresponding optimal measurement, and shortening the probe time of each probe. The optimal measurement is uniquely determined by the probe's initial state, and the precision limit obtained with the atom initially in the excited state is four times higher than that for an equal-weight superposition state. The estimation precision displays damped oscillatory behavior as the atom-boundary or atom-string separation increases. In the boundary case, the optimal polarization is always parallel to that boundary. In cosmic-string spacetime, by contrast, the optimal polarization is along the azimuthal direction around the string. The relative advantage of the other polarization directions depends on both the atom-string separation and the deficit angle. For small deficit angles and sufficiently large separations, radial polarization perpendicular to the string becomes less favorable than the other directions.
\end{abstract}


\maketitle

\section{Introduction}

In quantum metrology, unknown parameters can be estimated through measurements that are performed on the probe systems. The outcomes of the measurements follow a probability distribution which is determined by the unknown parameters. Using proper estimator, the unknown parameters can be estimated, and the uncertainty of the estimation satisfies the Cram\'{e}r-Rao bound~\cite{Helstrom,Holevo,Hubner,Braunstein}. The minimal uncertainty is determined by the times of probes as well as the Fisher information of the unknown parameter in each probe. With different models of the probe systems and different parameters to be estimated, many applications of quantum metrology have been done such as quantum frequency standards~\cite{Bollinger}, optimal quantum clock~\cite{Buzek}, measurement of gravity accelerations~\cite{Peters}, clock synchronization~\cite{Jozsa}, only to name a few. Through the system-environment interaction, factors of environment can be estimated, and the probe system is used as a detector.
One environment which no system can be isolated from is the vacuum that fluctuates all the time in quantum sense. Vacuum fluctuations can be modified by the space-time itself. Many efforts have been done to test the modification of vacuum fluctuations induced by the curved space-time. The test of quantum thermal effect induced by the un-inertial movements of observer draws great attention. Explicitly, the state of detector carried by uniformly accelerated observer will be thermalized, and the temperature of the thermalized state is determined by the acceleration~\cite{Unruh}. Restricted by the small acceleration that is obtained in usual experiment, the corresponding Unruh temperature becomes small quantity, and the test of such small quantity becomes a challenged issue.

In addition to the quantum thermal effect, quantum boundary effect is another kind of modification of quantum fluctuations induced by curved space-time. Quantum modes of fields can be reflected by the curved space-time itself.
Cosmic string space-time is one of the most simplest curved space-time, which modifies the vacuum fluctuations like a reflecting boundary does.
The cosmic strings appear as predictions of grand unification theories which are extended, one-dimensional, closed and infinite linear objects leading to topological defects of space-time. They emerge during the symmetry breaking phase transitions in the early universe~\cite{Kib,Vil} and become the probable sources of gravitational waves~\cite{Da,Br,Jac}, gamma-ray bursts~\cite{Ber}, and high-energy cosmic rays~\cite{Br09}. The idea of cosmic strings draws great attention in the context of the ``brane-world'' scenarios of the superstring theory~\cite{1,2,3,4,5,6}.
In cosmic string space-time, space-time is locally flat and what distinguishes it from a Minkowski spacetime is its nontrivial topology characterized by the deficit angle. Since the cosmic string only modifies the global space-time topology while leaving the local space-time flatness, it is pretty like what a conducting boundary does to a flat space-time. Modification of electromagnetic fluctuations induced by conducting boundary can be easily tested by polarizable atom that the evolution of the atom becomes position-dependent. Similarly, to test the modification of vacuum fluctuation induced by cosmic string, the probe polarizable atom can be used. A straightforward way to test whether the atomic state becomes position-dependent is to perform measurement on the probe atom, and repeat the process of preparation of initial atomic state, atom-field evolution and measurement several times. The position factor relative with the cosmic string can be estimated with limited uncertainty through maximum-likelihood estimation from the results of the measurements. Therefore, for given total time of all probes, by optimizing the preparation, evolution and measurements processes in each probe, the precision of estimation of the position factor will be enhanced, and the quantum boundary effects induced by the cosmic string will be tested with minimal uncertainty. Here we should note, the total probe time is the most crucial resource restricting the precision limit attainable in real experiments. A fixed total probe time can be distributed among a large number of probes. On the one hand, a shorter evolution time per probe reduces the Fisher information. On the other hand, it increases the number of repeated measurements (i.e., the number of probes), which in turn suppresses the estimation uncertainty. Therefore, for a fixed total probe time, it is the Fisher information per unit probe time, rather than the total Fisher information itself, that should be optimized to achieve the minimum estimation uncertainty. The optimal Fisher information, namely the quantum Fisher information, for several parameters including the initial state factors, the deficit angle factor, and the dynamics of an atom coupled to fluctuating quantum fields in cosmic string spacetime has been studied in Refs.~\cite{Cai,Zhou,Jin20,c1,c2,c3,c4}. In those works, the boundary-like effect leaves imprints on both the evolution of the atomic state and the estimation of the atomic state factor. However, to directly estimate the boundary-like effect itself, one must analyze how the probe system's state becomes position-dependent as the evolution time increases, as well as how to optimize the Fisher information per unit probe time specifically for the position factor. This is precisely the issue we will address in the present work. We begin by reviewing the evolution of a two-level atom and vacuum electromagnetic field, and to analyze how the atomic state is affected by the position factor.
\section{Optimal estimation of environmental factors through single two-level atom}
The total Hamiltonian of a two-level polarizable atom and electromagnetic field is given by
$
H=H_s+H_f+H_I
$.
Here $H_s={1\over
2}\,\hbar\omega_0\sigma_3$ denotes the Hamiltonian of
the atom with $\omega_0$ denoting the transition frequency. $H_f$ denotes the
Hamiltonian of the free electromagnetic field. $H_I$ denotes the interaction Hamiltonian
between the polarized atom and the electromagnetic field, and it is written in the multipolar
coupling scheme as
\begin{equation}\label{HI}
 H_I(\tau)=-e\textbf{r} \cdot
\textbf{E}(x(\tau))\;,
\end{equation}
with {\it e} denoting the electron
electric charge, $e\,\bf r$  denoting the atomic electric dipole moment, and
${\bf E}(x)$ denotes the electric field strength.
In the present work, the dynamics of the atomic state are analyzed in the frame of atom, and the field is assumed to be in vacuum state initially in such frame. So
the initial total density matrix is $\rho_{tot}(0)=\rho(0)\otimes|0\rangle\langle0|$ with $\rho(0)$ denoting the initial density matrix of atom and $|0\rangle$ denoting the vacuum state of field in the frame of atom. Under weak coupling, the reduced density matrix evolution follows the Kossakowski-Lindblad form~\cite{Lindblad,pr5}
\begin{equation}\label{master}
{\partial\rho(\tau)\over \partial \tau}= -\frac{I}{\hbar}\big[H_{\rm eff},\,
\rho(\tau)\big]
 + {\cal L}[\rho(\tau)]\ ,
\end{equation}
where $I$ denotes the imaginary unit, $\mathcal{L}[\rho]=\frac{1}{2}\sum_{i,j}a_{ij}(2\sigma_j\rho\sigma_i-\sigma_i\sigma_j\rho-\rho\sigma_i\sigma_j)$, and
the Kossakowski coefficients $a_{ij}=A\delta_{ij}-IB\epsilon_{ijk}\delta_{k3}-A\delta_{i3}\delta_{j3}$, with $A=\frac{1}{4}[\mathcal{G}(\omega_0)+\mathcal{G}(-\omega_0)]$ and $B=\frac{1}{4}[\mathcal{G}(\omega_0)-\mathcal{G}(-\omega_0)]$. Here
\begin{equation}
\mathcal{G}(\lambda)=\int_{-\infty}^{\infty}d\Delta\tau\,e^{I\lambda\Delta\tau}G^{+}(\Delta\tau),
\end{equation}
with $G^{+}(x-x')=\frac{e^2}{\hbar^2}\sum_{i,j}\langle+|r_i|-\rangle\langle-|r_j|+\rangle\langle0|E_i(x)E_j(x')|0\rangle$, and $|+\rangle$, $|-\rangle$ denoting the excited and ground state of the probe atom.
The effective Hamiltonian is $H_{\rm eff}=\frac{1}{2}\hbar\Omega\sigma_3$, with $\Omega=\omega_0+\frac{I}{2}[\mathcal{K}(-\omega_0)-\mathcal{K}(\omega_0)]$ and $\mathcal{K}(\lambda)=\frac{P}{\pi I}\int_{-\infty}^{\infty}d\omega\frac{\mathcal{G}(\omega)}{\omega-\lambda}$. The energy shift term in $H_{\rm eff}$ is small quantity, and is neglected in the following discussion.
The reduced density matrix can be written as $\rho=\frac{1}{2}(I+\mathbf{\omega}\cdot\mathbf{\sigma})$ with $I$ denoting the unit matrix, $\mathbf{\sigma}$ denoting the Pauli matrix and $\mathbf{\omega}$ denoting the Bloch vector.
For an initial state $\cos\frac{\theta}{2}|+\rangle+\sin\frac{\theta}{2}e^{I\phi}|-\rangle$, Eq.~(\ref{master}) is solved, and the Bloch vector evolves as
\begin{eqnarray}
\omega_1(\tau)&=&\sin\theta\cos(\Omega\tau+\phi)e^{-2A\tau},\nonumber\\
\omega_2(\tau)&=&\sin\theta\sin(\Omega\tau+\phi)e^{-2A\tau},\\
\omega_3(\tau)&=&\cos\theta e^{-4A\tau}-\frac{B}{A}(1-e^{-4A\tau}).\nonumber
\end{eqnarray}

The evolution of the state of the two-level atom is determined by factors $A$ and $B$, which are in relation with the field correlation function $\langle0|E_i(x)E_j(x')|0\rangle$. The distribution of field modes affect the evolution of the atom. Since the field is initially in vacuum state in the frame of the atom, spontaneous excitation will not occurs, and therefore, $\mathcal{G}(-\omega_0)=0$, and $A=B$. The distribution of field modes is determined by the structure of space-time. If the space-time is globally flat, applying the trajectory of the atom,
the electric-field two-point functions can be calculated,
and all coefficients of atomic evolution are calculated with
$
A=B=\frac{\gamma_0}{4}\;.
$
Here $\gamma_0=e^2|\langle -|{\bf
r}|+\rangle|^2\,\omega_0^3/3\pi\varepsilon_0\hbar c^3$ is the spontaneous emission rate in vacuum without boundaries.
Apparently, the evolution of the atom is independent of the position of the atom in the globally flat space-time. However, if the field is reflected by boundary or by the space-time itself, the field correlation function is modified, and therefore quantum boundary effect occurs, and the evolution factor $A$ may become position dependent. To estimate the boundary effect in such space-time, measurement is performed on the state of the atom.
Without loss of generality, the measurement is assumed in the base of
$\{\cos[\frac{\alpha}{2}+(-1)^i\frac{\pi}{2}]|+\rangle+e^{I\beta}\sin[\frac{\alpha}{2}+(-1)^i\frac{\pi}{2}]|-\rangle\}$, $(i=0,1)$.
Here $\alpha$ and $\beta$ are arbitrary weight and phase factor of the base of measurement, and $i$ denotes the result of measurement.
Therefore the probability with result $i$ becomes
\begin{equation}
P_i=\frac{1}{2}[1+(-1)^i\omega_1(\tau)\sin\alpha\cos\beta+(-1)^i\omega_2(\tau)\sin\alpha\sin\beta+(-1)^i\omega_3(\tau)\cos\alpha]\;.
\end{equation}
The above procedure including preparation of initial state of atom, atom-field evolution and measurement can be repeated $N$ times. By using maximum-likelihood estimation, the position factor $X$ can be estimated, and the uncertainty of the estimation satisfies~\cite{Helstrom,Holevo,Hubner,Braunstein}
\begin{equation}
Var (X)\geq\frac{1}{N F_X}
\end{equation}
with
\begin{equation}\label{F}
F_X=\sum_i\frac{(\partial_X P_i)^2}{P_i}=\frac{[\partial_X \omega_1(\tau)\sin\alpha\cos\beta+\partial_X \omega_2(\tau)\sin\alpha\sin\beta+\partial_X \omega_3(\tau)\cos\alpha]^2}{1-[\omega_1(\tau)\sin\alpha\cos\beta+\omega_2(\tau)\sin\alpha\sin\beta+\omega_3(\tau)\cos\alpha]^2}\;.
\end{equation}
We let $T$ denotes the total time of $N$ probes. So $N=\frac{T}{\tau}$. The minimal uncertainty of the estimation becomes
\begin{equation}
Var (X)\geq\frac{1}{N F_X}=\frac{1}{T F_X/\tau}\;
\end{equation}

\begin{figure}[htbp]
\centering
\includegraphics[height=2.1in,width=3.1in]{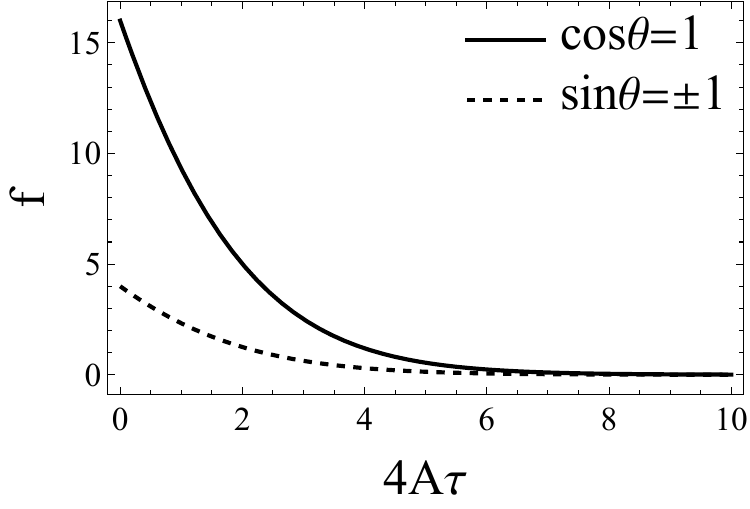}
\caption{ $f$ as  a function of $4A\tau$ with $\cos\theta=1$ and $\sin\theta=\pm1$.
}\label{p1}
\end{figure}

For given total probe time $T$, the precision of estimation is determined by $F_X/\tau$. The probe time $\tau$ of each probe, the initial state of probe atom, and the corresponding basis of measurement in the probe should be optimized to obtain the largest $F_X/\tau$.
For $\sin\theta=\pm1$, which corresponds to initial state $\frac{1}{\sqrt{2}}(|+\rangle\pm e^{I\phi}|-\rangle)$, the largest $F_X/\tau$ is obtained by choosing the basis of measurement with $|\sin\alpha|=1$, $\beta=\Omega\tau+\phi$. Therefore, according to Eq.~(\ref{F}),$F_X/\tau$ can be calculated as
\begin{equation}
F_X/\tau=f(4A\tau)|_{\sin\theta=\pm1}\times\frac{1}{4A} (\frac{\partial A}{\partial X})^2\;,
\end{equation}
with
\begin{equation}
f(4A\tau)|_{\sin\theta=\pm1}=\frac{4e^{-4A\tau}(4A\tau)}{1-e^{-4A\tau}}\;.\label{fs}
\end{equation}
While for $\cos\theta=1$, which corresponds to initial state $|+\rangle$, the largest $F_X/\tau$ is obtained by choosing measurement with $|\cos\alpha|=1$. Applying such state factors to Eq.~(\ref{F}),$F_X/\tau$ becomes
\begin{equation}
F_X/\tau=f(4A\tau)|_{\cos\theta=1}\times\frac{1}{4A} (\frac{\partial A}{\partial X})^2\;,
\end{equation}
with
\begin{equation}
f(4A\tau)|_{\cos\theta=1}=\frac{64e^{-2(4A\tau)}(4A\tau)}{1-(2e^{-4A\tau}-1)^2}\;.\label{fc}
\end{equation}
To optimize the probe time $\tau$ of each probe, $f$ in the two cases are plotted in function of $4A\tau$ in Fig.~(\ref{p1}) with the solid line denoting the case $\cos\theta=1$, and the dashed line denoting the case $\sin\theta=\pm1$.

Results show that $f$ decreases with the increasing of $\tau$ in both cases. The maximum of $f$ is reached when $4A\tau\rightarrow0$. As a result, the precision of estimation will be enhanced by reducing the probe time $\tau$ of each probe. When $4A\tau\rightarrow0$, $f(4A\tau)|_{\cos\theta=1}$ in Eq.~(\ref{fc}) approaches to $16$, and $f(4A\tau)|_{\sin\theta=\pm1}$ in Eq.~(\ref{fs}) approaches to $4$. Therefore, the maximum $F_X/\tau$ in case $\cos\theta=1$ becomes four times of that in case $\sin\theta=\pm1$. So $F_X/\tau$ reaches its maximum value
\begin{equation}\label{Ft}
F_X/\tau=\frac{4}{A} (\frac{\partial A}{\partial X})^2\;,
\end{equation}
when the initial state in each probe is chosen as the exited state, and the basis of measurement of each probe is constructed by the excited and ground state.
Here we should note, the Born--Markov approximation requires $\tau_B \ll \tau_R$ with $\tau_B$ denoting the bath correlation time (environmental memory time) and $\tau_R$ denoting the system relaxation time. The bath correlation time is determined by the inverse transition frequency of the atom  $\tau_B \sim 1/\omega_0$. For $\omega_0 \sim 10^{15}\,\mathrm{rad\,s^{-1}}$, this gives $\tau_B \sim 10^{-15}\,\mathrm{s}$, while the system relaxation time, i.e.\ the spontaneous emission lifetime, is $\tau_R = 1/\gamma_0 \sim 10^{-9}\,\mathrm{s}$. As shown in Fig.~\ref{p1}, when $4A\tau=0.1$, $f$ becomes $15.2133$, which is more than $95\%$ of its maximum value $16$ attained as $4A\tau\to0$.. Therefore, in real experiment, the evolution time in each probe can be chosen as $4A\tau=0.1$, which becomes $\gamma_0 \tau = 0.1$ in case of small deficit angle. Such protocol duration is also much larger than the bath correlation time, and therefore the evolution of probe atom is still within the Markovian regime.
In the following discussions, we use such settings.
\section{Optimal estimation of boundary effect in bounded flat space-time}
In the presence of infinite conducting reflecting boundary placed at $z=0$, the electric field correlation function $\langle0|E_i(x)E_j(x')|0\rangle$ for a static two-level atom is modified, and its Fourier transform can be calculated as~\cite{Yu12,Jin}:
\begin{equation}
{\cal G}(\lambda)=\sum_i\frac{e^2|\langle -|r_i|+\rangle|^2\lambda^3}{3\pi\varepsilon_0\hbar c^3}(1-f_i(\lambda,z))\theta(\lambda)\;,
\end{equation}
with
\begin{eqnarray}
f_x(\lambda,z)=&&f_y(\lambda,z)=\frac{3c^3}{16\lambda^3z^3}\bigg[\frac{2\lambda z}{c}\cos\frac{2\lambda z}{c}+\bigg(\frac{4\lambda^2z^2}{c^2}-1\bigg)\sin\frac{2\lambda z}{c}\bigg]\;,\nonumber\\
&&f_z(\lambda,z)=\frac{3c^3}{8\lambda^3 z^3}
\left[
\frac{2\lambda z}{c}\cos\frac{2\lambda z}{c}
-\sin\frac{2\lambda z}{c}
\right].
\end{eqnarray}
Therefore
$
A=B=\frac{\gamma_0}{4}(1-\sum_i\alpha_if_i(\omega_0,z))\;.
$
%~\footnote{The divergence in the integration in Eq. (\ref{lm}) can be treated by introducing a cut-off factor $\omega_M$.}
Here
$\alpha_i=|\langle
-|r_i|+\rangle|^2/|\langle -|{\bf r}|+\rangle|^2\,.$ $\alpha_i$, which are usually dubbed as the relative polarizability satisfying $\sum_i\alpha_i=1$.
Apparently, the polarization direction of the atom affects the evolution of the atom, and therefore affects the probability of the measurements. For $z\rightarrow\infty$, $f_i(\omega_0,z)\rightarrow0$, the present case returns back to the unbounded Minkowski vacuum case. While for $z\rightarrow0$, $1-f_x=1-f_y=\frac{4(\omega_0z)^2}{5}$, and $1-f_z=2$. Therefore, only the state of atom with parallel polarization becomes position-dependent.
After $N$ probes, using maximum-likelihood estimation, position factor $z$ can be estimated. For a fixed total time $T$ of all probes, the precision limit of the estimation depends on $F_z/\tau$. $F_z/\tau$ for both the parallel and vertical cases are plotted in function of $z$.
\begin{figure}[htbp]
\centering
\includegraphics[height=2.1in,width=3.1in]{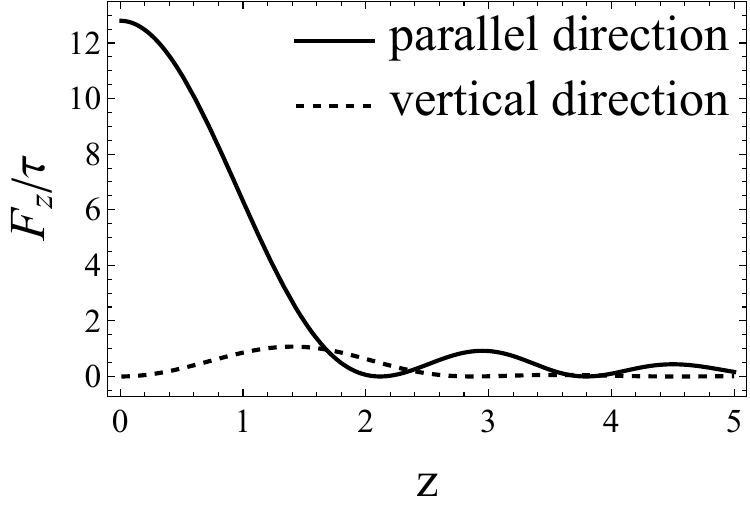}
\includegraphics[height=2.1in,width=3.1in]{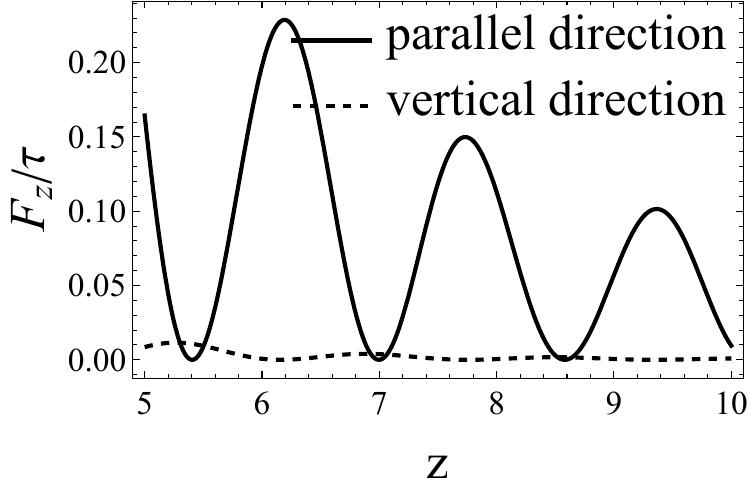}
\caption{ $F_z/\tau$ as  a function of $z$ for in cases of parallel and vertical polarization respectively. Here $F_z/\tau$ is in the unit of $\frac{\omega_0^2\gamma_0}{4c^2}$, and $z$ is in the unit of $c/\omega_0$.
}\label{F1}
\end{figure}
Results show that in both small separation area with $0<z<5c/\omega_0$ and long separation are $z>5c/\omega_0$, the averaged precision in parallel case is larger than that in vertical case. Therefore, the quantum boundary effect in flat space time with reflecting boundary can be estimated with larger precision by using probe atoms with the direction of polarization being parallel with the boundary.
\section{Optimal estimation of boundary effect in cosmic string space-time}
A static, straight cosmic string is assumed to be lying along the $z$-direction. So the metric in the cylindrical coordinate system $(t,\rho,\varphi,z)$ is expressed as
$
d s^2=c^2d t^2-d\rho^2-\rho^2d\varphi^2-d z^2\;,
$
where $0\leq\varphi<\frac{2\pi}{\nu}$, $\nu=(1-4G\mu)^{-1}$ with $G$ and $\mu$ being the Newton's constant and the mass per unit
length of the string respectively. The deficit angle $\Delta$ becomes $\Delta= 2\pi(1-1/\nu)=8\pi G\mu$. We should note $\Delta\rightarrow0$ corresponds to $\nu-1\rightarrow0$.
In such space-time background,
the electric field correlation function $\langle0|E_i(x)E_j(x')|0\rangle$ for a static two-level atom is calculated, and the Fourier transform of the correlation functions is expressed as~\cite{Cai,Zhou,Jin20,c1,c2,c3,c4}
\begin{equation}
{\cal G}(\lambda)=\sum_i\frac{e^2|\langle -|r_i|+\rangle|^2\lambda^3}{3\pi\varepsilon_0\hbar c^3}f_i(\lambda,\rho,\nu)\theta(\lambda)\;,
\end{equation}
with
\begin{eqnarray}
f_\rho(\lambda,\rho,\nu)&=&\frac{3\nu}{4}\sum_m\int_0^{1}dt\frac{t}{\sqrt{1-t^2}}[(2-t^2)J^2_{|\nu m+1|}(\lambda\rho t)\nonumber\\
&&+t^2J_{|\nu m|+1}(\lambda\rho t)J_{|\nu m|-1}(\lambda\rho t)]\;,\nonumber\\
f_\varphi(\lambda,\rho,\nu)&=&\frac{3\nu}{4}\sum_m\int_0^{1}dt\frac{t}{\sqrt{1-t^2}}[(2-t^2)J^2_{|\nu m+1|}(\lambda\rho t)\nonumber\\
&&-t^2J_{|\nu m|+1}(\lambda\rho t)J_{|\nu m|-1}(\lambda\rho t)]\;,\nonumber\\
f_z(\lambda,\rho,\nu)&=&\frac{3\nu}{2}\sum_m\int_0^{1}dt\frac{t^3}{\sqrt{1-t^2}}J^2_{|\nu m|}(\lambda\rho t)\;.\label{definition f3}
\end{eqnarray}
Here the symbol $J$ denotes BesselJ function.
Therefore
$
A=B=\frac{\gamma_0}{4}[\sum_i\alpha_if_i(\omega_0,\rho,\nu)]\;.
$
%~\footnote{The divergence in the integration in Eq. (\ref{lm}) can be treated by introducing a cut-off factor $\omega_M$.}
Similarly, the polarization direction of the atom affects the evolution of the atom, and therefore affects the probability of the measurements. For $\rho\rightarrow\infty$, $f_i\rightarrow1$, the present case returns back to the unbounded Minkowski vacuum case. While for $\rho\rightarrow0$, $f_\rho\approx f_\varphi\approx\frac{3\nu^2(\nu+1)}{\Gamma[2\nu+2]}(\omega_0\rho)^{2(\nu-1)}$, $f_z\approx\nu$. The evolution of probe atom becomes similar to that in reflected boundary case especially for $\nu=2$.
Therefore, we first consider the case that the topological deficit of angle is significant with $\nu=2$. The quantum boundary effect becomes significant in such case. Using maximum-likelihood estimation, position factor $\rho$ can be estimated. For a fixed total time $T$ of all probes, the precision limit of the estimation depends on $F_\rho/\tau$. In case $\nu=2$, $F_\rho/\tau$ for the three polarization directions cases are plotted in function of $\rho$.
\begin{figure}[htbp]
\centering
\includegraphics[height=2.1in,width=3.1in]{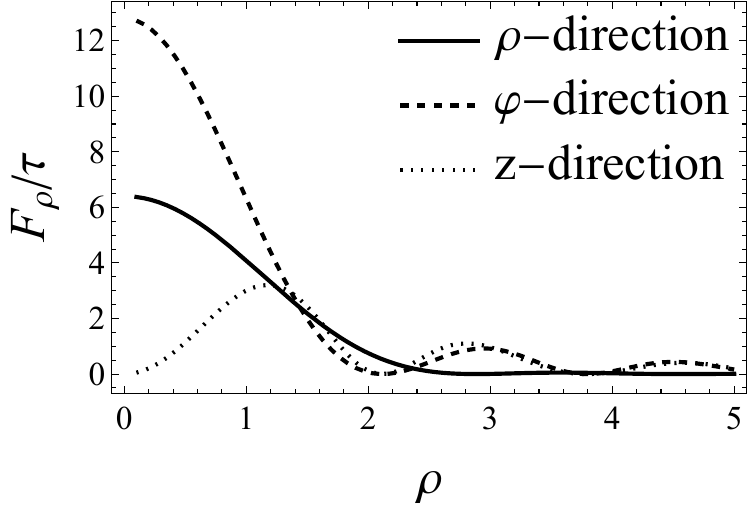}
\includegraphics[height=2.1in,width=3.1in]{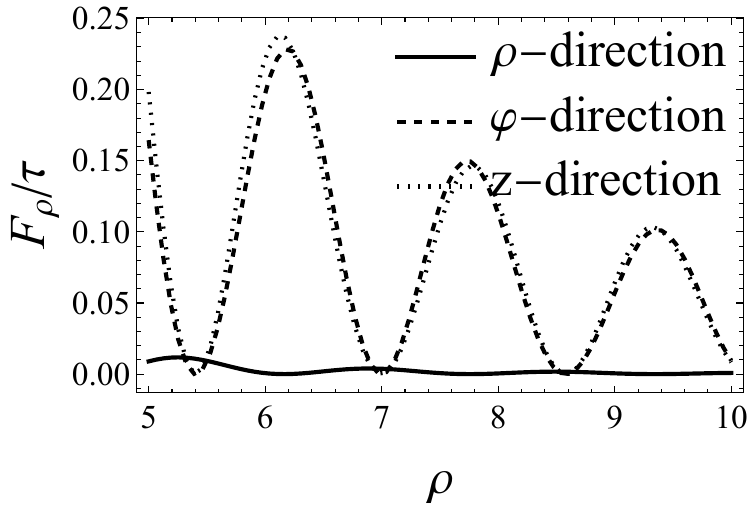}
\caption{ $F_\rho/\tau$ as a function of $\rho$ in cases with $\rho$, $\varphi$ and $z$ polarization direction when $\nu=2$. Here $F_\rho/\tau$ is in the unit of $\frac{\omega_0^2\gamma_0}{4c^2}$, and $\rho$ is in the unit of $c/\omega_0$.
}\label{F2}
\end{figure}
Results show that in small separation area with $0<\rho<5c/\omega_0$, the averaged precision in $\phi$-direction is larger than those in other cases, and that in $z$-direction case is the smallest. While in long separation area $\rho>5c/\omega_0$, the averaged precision in $\rho$-direction case becomes the smallest. The averaged precision in $z$-direction and $\phi$-direction cases become almost the same. Therefore, the quantum boundary effect in cosmic string space time can be estimated with larger precision by using $\phi$-direction probe atom in both small separation area and large separation area.

\begin{figure}[htbp]
\centering
\subfloat[$\nu-1=0.5$]{%
    \includegraphics[height=1.7in, width=2.1in]{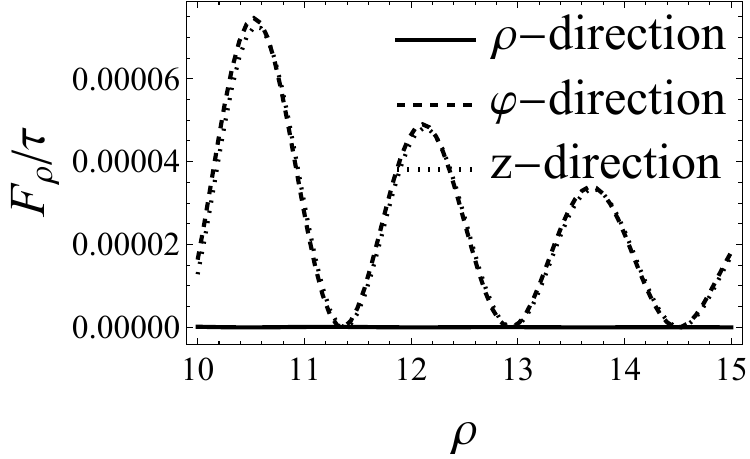}%
}%
\hfill
\subfloat[$\nu-1=0.2$]{%
    \includegraphics[height=1.7in, width=2.1in]{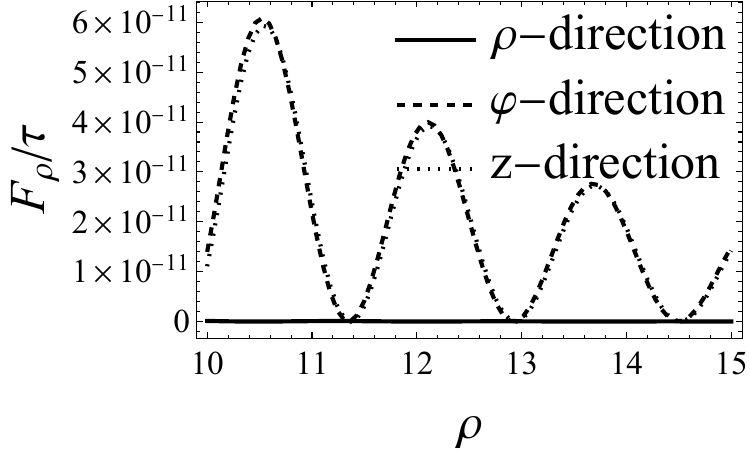}%
}%
\subfloat[$\nu-1=0.1$]{%
    \includegraphics[height=1.7in, width=2.1in]{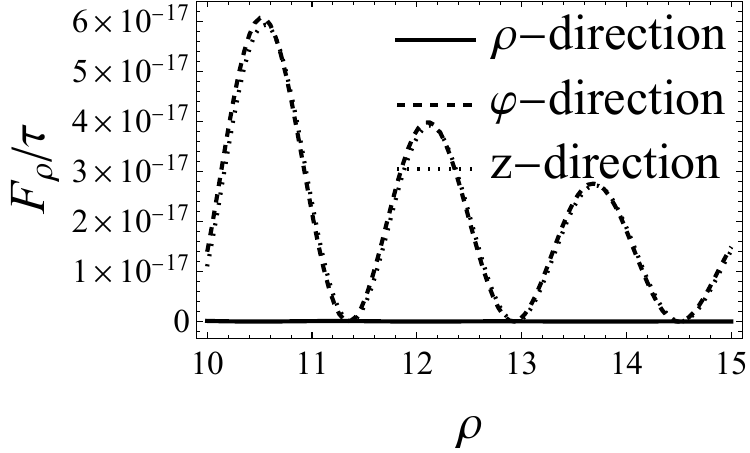}%
}%
\caption{$F_\rho/\tau$ as a function of $\rho$ in cases with $\rho$, $\varphi$ and $z$ polarization direction when $\nu-1=10^{-1},10^{-4},10^{-7}$ respectively. Here $F_\rho/\tau$ is in the unit of $\frac{\omega_0^2\gamma_0}{4c^2}$, and $\rho$ is in the unit of $c/\omega_0$.}
\label{c1}
\end{figure}

However, the case $\nu = 2$ corresponds to $\Delta = \pi$, which is far outside the experimentally allowed range. It is included only as an idealized illustrative limit. In practical, the deficit angle $\nu-1\rightarrow0$~\cite{Vi,Dvo}, and $\rho$ should be considered with large value.
An optimistic estimate of the linear mass density of the cosmic string formed via the Kibble mechanism during a phase transition in the early Universe is $G\mu\sim10^{-7}$~\cite{Vi,Dvo}. This means that the parameter $\nu$ differs from unity at the seventh decimal place $\nu-1=10^{-7}$. In Fig.~\ref{c1}, we analyze the small-deficit-angle regime using $\nu-1=10^{-1}$, $10^{-4}$, and $10^{-7}$. These values decrease gradually toward $\nu-1\to0$. 
Results show that with the decrease of $\nu-1$, $F_\rho/\tau$ in $z$-direction case and in $\varphi$-direction case become almost identical, and the value of $F_\rho/\tau$ in such cases is much larger than that in $\rho$-direction case. Though the value of $F_\rho/\tau$ becomes small quantity with $\nu-1\rightarrow0$ at large separation area, the precision of estimation can be enhanced by increasing the total probe time $T$.

\section{Conclusion}
In summary, quantum boundary effects can be estimated by performing $N$ sequential measurements on a single probe atom. For a fixed total time $T$ of all probes, the measurement basis, the initial state of each atom, the interaction time of each probe, and the polarization direction of the atom are optimized to minimize the estimation uncertainty in both scenarios where vacuum fluctuations are modified either by a reflecting boundary or by a cosmic string. In both cases, the optimal measurement basis is constructed from the excited and ground states in the sequential measurement process, with each probe initialized in the excited state. However, the optimal polarization configurations differ in the two scenarios.

In the reflecting-boundary case, polarization parallel to the boundary always outperforms the perpendicular polarization. In contrast, in the cosmic-string space-time, the azimuthal direction around the string becomes the optimal polarization. For a large deficit angle case with $\nu = 2$, and in the short-separation regime $0 < \rho < 5c/\omega_0$, polarization along the $\phi$-direction (i.e., azimuthal direction around the string) yields the minimal estimation uncertainty and thus becomes optimal. In the long-separation regime $\rho > 5c/\omega_0$, the precision limit for polarization along the $z$-direction (i.e., the direction of the cosmic string) becomes identical for the $\phi$-direction, while both significantly exceed that obtained for polarization along the $\rho$-direction (i.e., the radial atom-string separation direction). For the physically relevant cases of small deficit angles toward $\nu-1\rightarrow0$ and large atom-string separation, similar to the long-separation regime in the large deficit angle case, $z$-direction and $\varphi$-direction become the optimal polarization directions.

\begin{acknowledgments}
This work was supported by the National Natural Science Foundation of China under Grants No. 12165003, the special funding of talent program in Guizhou province[GCC[2023]005].
\end{acknowledgments}

\end{document}